\documentclass[preprintnumbers,amsmath,amssymb]{revtex4}
\usepackage{amsfonts}
\usepackage{graphics}
\usepackage{epsfig}
\usepackage{url}
\usepackage{multirow}
\newcommand {\be}{\begin{equation}}
\newcommand {\ee}{\end{equation}}
\newcommand {\ba}{\begin{eqnarray}}
\newcommand {\ea}{\end{eqnarray}}

\newcommand{\tr}{\ensuremath{\mathrm{tr}}}

\newcommand{\I}{\ensuremath{\mathrm{i}}}

\newcommand{\MeV}{\ensuremath{\,\mathrm{MeV}}}
\newcommand{\GeV}{\ensuremath{\,\mathrm{GeV}}}

\newcommand{\braket}[1]{\ensuremath{\left<#1\right>}}
\newcommand{\hc}{\ensuremath{\text{h.c.}}}

\newcommand{\SU}[1]{\ensuremath{\mathrm{SU}(#1)}}
\newcommand{\U}[1]{\ensuremath{\mathrm{U}(#1)}}

\begin{document}
\title{Strategies to link tiny neutrino masses with huge missing mass of
the Universe }
\author{Y. Farzan }

\affiliation{School of physics, Institute for Research in
Fundamental Sciences (IPM), P.O. Box 19395-5531, Tehran, IRAN}
\email{yasaman@theory.ipm.ac.ir}
\begin{abstract}
With the start of the LHC, interest in electroweak scale models
for the neutrino mass has grown. In this letter, we review two
specific models that simultaneously explain neutrino masses and
provide a suitable DM candidate. We discuss the implications of
these models for various observations and experiments including
the LHC, Lepton Flavor Violating (LFV) rare decays, direct and
indirect dark matter searches and Kaon decay.

\end{abstract}
 \maketitle
\section{Introduction}
Although the Standard Model (SM) of particle physics has been
triumphant in explaining the observations in high energy
accelerators, it has failed to accommodate the Lepton Flavor
Violation (LFV) in the neutrino oscillation. Moreover, there is no
suitable candidate for dark matter in the SM. Extensive literature
exists on models explaining each of these phenomena. For example,
the conventional type I seesaw mechanism with very heavy
right-handed neutrinos explains the small neutrino masses.
However, this model does not provide a candidate for Dark Matter
(DM). Moreover, since the scale of new physics in this scenario
({\it i.e.,} the masses of the right-handed neutrinos) is much
higher than the reach of the LHC or any other man-made accelerator
in the foreseeable future, directly testing this scenario is a
dream that may not ever come true.

 Testable models that can
simultaneously explain neutrino masses and dark matter have been
in the center of attention in recent years. To a great extent,
this attention owes to the start of the LHC and a host of direct
and indirect DM search experiments that are either in the process
of data taking or planned to do so in the near future. To explain
neutrino oscillation, the model should include sources for LFV. If
the mass scale of the new particles is low, we in general expect
rather large LFV decay rates, Br($\mu \to e \gamma$), Br($\tau \to
e \gamma$) and Br($\tau \to \mu \gamma$), that can be in principle
observed at the experiments searching for these decays. With the
MEG experiment taking data  \cite{MEG} and the superflavor project
under consideration, this possibility is becoming more exciting.

In this article, we review two specific models that can
simultaneously explain neutrino masses and provide a DM candidate.
Both models are based on a $Z_2$-symmetry under which all SM
particles are even and all new particles are odd. As a result, the
lightest new particle is stable against decay and provides a
suitable DM candidate. In both these models, right-handed
neutrinos are added to the SM; however, the $Z_2$ symmetry
prevents a Dirac mass term for neutrinos. Neutrinos acquire
Majorana mass terms at one-loop level.

One of these models, the so-called SLIM model, contains a low
energy sector with masses less than $O(10)$ MeV
\cite{Boehm,myModel}. The low energy sector can show up at various
observable phenomena like rare Kaon decay or supernovae. As we
shall see, the upper bounds on the masses of these low energy
sector particles and lower bounds on the couplings provide a way
to test the SLIM model. The second model, the so-called AMEND
model \cite{AMEND}, does not contain such a low energy sector;
however, it can lead to interesting phenomenology at direct
searches for DM and at the LHC.

The paper is organized as follows: In section \ref{SLimplication},
we  review the structure of the low energy SLIM scenario and its
implication for low energy observations. In section \ref{UV}, we
explain how the low energy effective SLIM scenario can be embedded
within an electroweak symmetric model. In section \ref{imp}, we
review the prediction of this model for various observations such
as DM annihilation into electron-positron or photon pairs,
self-interaction of DM and LFV rare decays. In section \ref{LHC},
we discuss the possibility of discovering the model at the LHC. In
section \ref{mend}, we first describe  the structure of AMEND. We
then review the neutrino mass matrix, LFV rare decay, DM
annihilation and direct and indirect DM searches, electroweak
precision test and signatures at colliders within the context of
this model. Results are summarized in section \ref{con}.

\section{The real SLIM scenario and its implications \label{SLimplication}}
In this section, we review the so-called SLIM scenario which has
been first introduced in \cite{Boehm}. SLIM is the abbreviation of
Scalar as LIght as Mev which describes the characteristics of the
DM candidate within this scenario. The scenario is quite
minimalistic and adds only a scalar, $\delta$ and two (or more)
right-handed Majorana neutrinos, $N_i$. A $Z_2$ symmetry is
imposed on the scenario under which $\delta$ and $N_i$ are odd but
the SM particles are all even. As discussed in \cite{Boehm}, the
scenario can be realized both for real and complex $\delta$. We
will however concentrate on the real SLIM scenario. $\delta$,
which is called  SLIM, is lighter than the other $Z_2$ odd
particles and as a result is stable and can play the role of the
DM candidate.

\begin{figure} \label{oneLOOP}
\begin{center}
\includegraphics[width=6 cm,bb=54 120 337 277, clip=true]{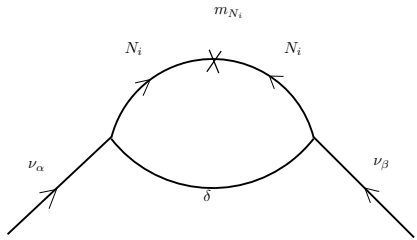}
\end{center}
 \caption{One-loop contribution to the neutrino mass}

\end{figure}

 The scenario is based on the following
Lagrangian: \be \label{Lag}\mathcal{L}=m_\delta^2{\delta^2 \over
2}+\left( g_{i \alpha}  \bar{N}_i \nu_\alpha\delta+{m_{N_i} \over
2}\bar{N_i^c} N_i +{\rm H.c.}\right).\ee Of course, if we assume
neither $N_i$ nor $\delta$ have electroweak interactions, the
$g_{i \alpha}$ coupling  in Eq. (\ref{Lag}) will not be symmetric
under SU(2)$\times$U(1). As a result, it can be only effective.
That is, the scenario should be augmented to become symmetric
under SU(2)$\times$U(1) at higher energies. This step will be
taken in the next section. In this section, we are concerned about
the low energy implications of the scenario. Notice that the
Majorana mass term for $N_i$ violate lepton number and the
coupling $g_{i \alpha}$ is a source of LFV. As shown \cite{Boehm},
this Lagrangian through the one-loop diagram shown in
Fig.~\ref{oneLOOP} leads to the Majorana mass for active neutrinos
\begin{equation}
\label{flavoredmass} (m_\nu)_{\alpha \beta} =\sum_i \frac{g_{i
\alpha}g_{i \beta}}{16
  \pi^2}m_{N_i}\left(\log
  \frac{\Lambda^2}{m_{N_i}^2}-\frac{m_\delta^2}{m_{N_i}^2-m_\delta^2}\log
\frac{m_{N_i}^2}{m_{\delta}^2} \right),
\end{equation}
where $\Lambda$ is the ultraviolet cutoff of the effective
coupling which is the electroweak scale. Let us parameterize the
neutrino mass matrix as follows \be \label{parNuM} m_{\nu}=U \cdot
{\rm Diag}[m_1,m_2 e^{2i \gamma_2},m_3e^{2i\gamma_3}] U^T \ .\ee
Combining (\ref{flavoredmass},\ref{parNuM}), we find
\begin{equation}
  \label{PhiRealComplex} g={\rm Diag}(X_1,..., X_n)\cdot
  O \cdot {\rm Diag}(\sqrt{m_1},\sqrt{m_2}e^{i \gamma_2},\sqrt{m_3}e^{i
    \gamma_3}) U^T \ ,
\end{equation}
where  $n$ is the number of Majorana neutrinos  and $O$ is an
arbitrary $n \times 3$ matrix that satisfies $O^T \cdot O={\rm
Diag}(1,1,1)$. Finally,
\begin{equation}
\label{real-Xi} X_i=4\pi
\left(\frac{1}{m_{N_i}}\right)^{1/2}\left( \log
\frac{\Lambda^2}{m_{N_i}^2} -{m_\delta^2 \over
m_{N_i}^2-m_\delta^2} \log \frac{m_{N_i}^2}{m_\delta^2}
\right)^{-1/2}.
\end{equation}
As is well-known, within thermal production scenario, the dark
matter energy budget of the universe is almost  independent of
mass and is determined by the DM annihilation cross section. From
this observation, the DM annihilation cross section is determined:
 $\langle \sigma v \rangle\sim 10^{-26}~{\rm cm}^3{\rm sec}^{-1}$. Within the SLIM scenario the main
annihilation modes are $\delta \delta \to \nu_\alpha \nu_\beta,
\bar\nu_\alpha \bar\nu_\beta$. Notice that these processes violate
lepton number by two units. Indeed, these processes proceed
through $t$-channel diagrams with a lepton number violating
right-handed neutrino  propagator proportional to $m_{N_i}$:
\begin{equation} \label{sigma}
\langle \sigma(\delta \delta \to \nu_\alpha \nu_\beta ) v
\rangle=\langle \sigma(\delta \delta \to \bar{\nu}_\alpha
\bar{\nu}_\beta ) v \rangle=\frac{1}{4\pi}\left|\sum_i {g_{i
\alpha } g_{i\beta } m_{N_i}\over m_\delta^2+m_{N_i}^2}\right|^2.
\end{equation}
Taking $\Lambda \sim 200$~GeV, $0.01~{\rm eV}<m_\nu<1~{\rm eV}$
and $\langle\sigma(\delta \delta \to {\rm everything})v
\rangle\sim 10^{-26}~{\rm cm}^3 {\rm sec}^{-1}$, from
Eqs.~(\ref{flavoredmass},\ref{sigma}), we obtain \be O(1)~{\rm
MeV}\stackrel{<}{\sim} m_{N_1}\stackrel{<}{\sim}10~{\rm MeV} \ \ \
\ \ \ {\rm and} \ \ \ \ \ \ \ 3 \times 10^{-4}\stackrel{<}{\sim}
g_{1 \alpha}\stackrel{<}{\sim}10^{-3} \ , \ee where $N_1$ is the
lightest right-handed neutrino whose propagator dominates the DM
annihilation. Remember that $\delta$ is taken to be lighter than
$N_i$ so the DM candidate within this scenario is lighter than
10~MeV. This argument does not set a lower bound on $m_\delta$;
however from primordial nucleosynthesis a lower bound of $O(1)$
MeV is derived \cite{serpicoraffelt}. That is within this
scenario, the DM mass is in the MeV range. This is the rationale
for naming $\delta$ as Scalar as LIght as Mev, SLIM.

Although in this scenario DM pairs mainly annihilate to neutrino
or antineutrino pairs, the energy of the produced neutrinos
($E_\nu\simeq m_\delta <10$~MeV) will be too low to be detectable
at neutrino telescopes such as ICECUBE. In principle, as shown in
\cite{silvia}, neutrinos produced by annihilation of 20-30 MeV DM
candidates in dark halo can be observed by future large neutrino
detectors such as LENA \cite{LENA}.  However, for the mass range
relevant for real SLIM, we do not expect a sizeable signal
\cite{Farzan}.

The most interesting feature of this scenario is that there is a
lower bound on the coupling of the new sector to neutrino and an
upper bound on their masses. This means by collecting enough data
in the low energy observations that involve neutrinos, this model
can be eventually tested. This feature and its phenomenological
implications have been elaborated on in \cite{Farzan}. Consider
the decay $A\to B+\nu$ where $A$ and $B$ can be detected but $\nu$
appears as missing energy. If the mass difference between $A$ and
$B$ is less than the sum of the masses of $\delta$ and $N_i$,
there will be another contribution to the missing energy signal:
$$A \to B N_i \delta$$ where  both $\delta$ and $N_i$, like $\nu$,
escape detection. $N_i$ eventually decay into $\nu\delta$. Thus,
by studying the decay mode $A\to B+{\rm missing ~energy}$,
information on the parameters of the scenario can be deduced.
{Similar analysis has extensively been carried out (see
Refs.~\cite{Britton:1993cj,Barger:1981vd,Gelmini:1982rr,Lessa}) in
the case of the Majoron couplings to neutrinos.} In particular,
consider the decay of $K^+$ to the charged leptons. By comparing
${\Gamma(K^+\to e^+ +{\rm missing~ energy}) /\Gamma(K^+\to \mu^+
+{\rm missing~ energy})}$ with the SM prediction, the coupling
$g_{i \alpha}$ can be constrained: \be \label{universality}
{\Gamma(K^+\to e^+ +{\rm missing~ energy}) \over\Gamma(K^+\to
\mu^+ +{\rm missing~ energy})}={ \Gamma_{SM}(K^+ \to e^+ \nu_e)
+\sum_i\Gamma(K^+ \to e^+ N_i \delta) \over \Gamma_{SM}(K^+ \to
\mu^+ \nu_e) +\sum_i\Gamma(K^+ \to \mu^+ N_i \delta)}\ee
$$ \simeq {
\Gamma_{SM}(K^+ \to e^+ \nu_e) +\sum_i\Gamma(K^+ \to e^+ N_i
\delta) \over \Gamma_{SM}(K^+ \to \mu^+ \nu_e)}.$$ In the last
line, we have taken
$$\Gamma(K^+\to e^+ N_i \delta)\ll \Gamma(K^+\to \mu^+ N_i
\delta)\times {\Gamma(K^+ \to e^+ \nu)\over \Gamma(K^+ \to \mu^+
\nu)}\ .$$ Considering the fact that $\Gamma(K^+ \to e^+
\nu)/\Gamma(K^+ \to \mu^+ \nu)\sim (m_e/m_\mu)^2\ll 1$, this
assumption is justified. The recent bounds on the ratio
${\Gamma(K^+\to e^+ +{\rm missing~ energy}) /\Gamma(K^+\to \mu^+
+{\rm missing~ energy})}$ from KLOE~\cite{Kloe} yields
$$ \sum_i |g_{ie}|^2<10^{-5}$$ where the sum runs over $N_i$
lighter than $K^+$.

The spectrum of the charged lepton in the two-body decay $K^+ \to
\ell^+ +\nu$ will be of course different from that in the three
body decay $K^+ \to \ell^+ +N_i+\delta$. Thus, by studying the
spectrum of $\mu^+$ in $K^+ \to \mu^+{\rm missing~ energy}$,
information can be derived on $\sum_i|g_{i \alpha}|^2$. The
analysis based on 1973 LBL Bevatron data \cite{73} gives
\cite{Farzan}
$$ \sum_i|g_{i \mu}|^2<9 \times 10^{-5}\ . $$
A more thorough investigation of $K^+ \to \mu^+ N_i \delta$ can be
performed with the present KLOE data as well as with the upcoming
NA62 results \cite{NA62}.

Another situation where the SLIM scenario can show up is the
supernova explosion during which a neutrino gas of temperature of
$T \sim  {\rm few}~10$~MeV is formed inside the supernova core.
Since the SLIM particles  are relatively light and are coupled to
neutrinos, they can be produced at the supernova explosion. The
produced $\delta$ can interact  with neutrinos in that environment
with a cross section given by \cite{Farzan} $$\sigma(\delta
N_i\to\delta N_i)\sim {g^4 T^2 \over 4\pi (T^2 +m_{N_i}^2)^2} .$$
Taking $T \sim {\rm few}\times 10 $ MeV, we find that the mean
free path of the SLIM particles is $(\sigma n_\nu)^{-1}\sim 10 $~
cm which is far shorter than the supernova core. As a result, the
SLIM will be trapped inside the core. The energy transferred by
diffuse out of the SLIM particles can be tolerated within the
present uncertainties of supernova model \cite{Boehm}. In case of
supernova explosions in the future, the scenario might be tested
by studying the neutrino energy spectra \cite{Boehm}.

\section{Ultraviolet completion of the SLIM scenario \label{UV}}
In the previous section, we introduced the low energy SLIM
scenario based on the effective Lagrangian in Eq.~(\ref{Lag}) and
briefly discussed its implications on the relevant low energy
phenomena. The Lagrangian in Eq.~(\ref{Lag}) has to be embedded
within a SU(2)$\times$U(1) symmetric model. In \cite{Boehm},
several ideas for the ultraviolet completion of the SLIM scenario
have been suggested. In \cite{myModel} a minimalistic model have
been introduced that embeds the SLIM scenario. In this section, we
review this model and in the next section we discuss its
implications.

The model presented in \cite{myModel} is quite minimalistic and is
composed of (1) an electroweak singlet, $\eta$; (2) two Majorana
right-handed neutrinos, $N_{i}$ and (3) an electroweak doublet
with nonzero hypercharge, $\Phi^T=[\phi^0 \ \phi^-]$ where
$\phi^0\equiv (\phi_1+i\phi_2)/\sqrt{2}$ with real $\phi_1$ and
$\phi_2$. As emphasized before,  all these new particles are odd
under the $Z_2$ symmetry. Imposing the $Z_2$ symmetry, the most
general $Z_2$ even renormalizable Lagrangian involving only the
scalars will be of form
\begin{align}
\mathcal{L}=&-m_\Phi^2 \Phi^\dagger \cdot \Phi -{m_s^2\over 2}
\eta^2
- (m_{\eta \Phi} \eta(H^T (i\sigma_2) \Phi)+{\rm H.c.}) \nonumber \\
~&-{\lambda_1 } |H^T (i\sigma_2) \Phi|^2-{\rm Re}[\lambda_2 (H^T
(i\sigma_2) \Phi)^2]-\lambda_3 \eta^2 H^\dagger H-\lambda_4
\Phi^\dagger \cdot \Phi H^\dagger \cdot H
\nonumber \\
~& -{\lambda'_1 \over 2} (\Phi^\dagger \cdot
\Phi)^2-{\lambda'_2\over 2} \eta^4-\lambda'_3\eta^2 \Phi^\dagger
\cdot \Phi \nonumber \\
~& -m_H^2 H^\dagger\cdot H- {\lambda \over 2}(H^\dagger \cdot
H)^2\ . \label{MainLag}
\end{align}
Positivity of the potential at infinity puts constraints on the
couplings \cite{G-H},
$$\lambda'_1,\lambda'_2>0, \
\lambda'_3>-(\lambda'_1\lambda'_2)^{1/2}, \
\lambda_3>-(\lambda\lambda'_2)^{1/2}$$ and $$
\lambda_1-|\lambda_2|+\lambda_4 > -(\lambda\lambda'_1)^{1/2}\ .
$$ Phases of $\lambda_2$ and $m_{\eta \Phi}$ are sources of
CP-violation. For simplicity, we impose CP-symmetry on the
Lagrangian in Eq.~(\ref{MainLag}) which makes all the parameters
in Eq.~(\ref{MainLag}) real.

Setting $H^T=(0 \ v_H/\sqrt{2})$, the mass terms will be of form
\begin{align}
\mathcal{L}_m&=-m_{\phi^-}^2|\phi^-|^2- \frac{m_{\phi_2}^2}{2}\phi_2^2 \nonumber\\
&-\frac{m_\eta^2}{2}\eta^2-\frac{m_{\phi_1}^2}{2} \phi_1^2
-m_{\eta \Phi}v_H \phi_1 \eta
\end{align} where
\begin{align}
m_{\phi^-}^2=&
m_\Phi^2+\lambda_4\frac{v_H^2}{2}\label{phi-}\\
m_\eta^2=&m_s^2+\lambda_3\frac{v_H^2}{2}\\
m_{\phi_1}^2 =& m_\Phi^2+\lambda_1\frac{v_H^2}{2}+
\lambda_2\frac{v_H^2}{2} \\
m_{\phi_2}^2
=&m_\Phi^2+\lambda_1\frac{v_H^2}{2}-\lambda_2\frac{v_H^2}{2}\ .
\label{phi2}
\end{align}
The parameters are taken in a range that neither of the scalars,
except the SM Higgs, develops a vacuum expectation value.
$\phi^-$, being a charged particle should be heavier than $\sim
80$~GeV to avoid the direct search bounds \cite{chargedHiggs}.
Notice that while $\phi_1$ mixes with $\eta$ through the $m_{\eta
\Phi}$ term, there is no such a mixing between $\phi_2$ and
$\eta$. Had we taken $m_{\eta \Phi}$ complex, the mixing term
would be
$$\Re[m_{\eta \Phi}] v_H \phi_1 \eta+\Im[m_{\eta \Phi}] v_H \phi_2
\eta\ .$$ However as we discussed above, we take the Lagrangian to
be CP-symmetric so $\phi_2$ is a mass eigenstate itself. The other
neutral mass eigenstates
 are $\delta_1$ and $\delta_2$ defined as follows:
\ba \label{mixing-ETA-PHI1} \left[
\begin{matrix}\delta_1 \cr \delta_2 \end{matrix}\right]=
\left[
\begin{matrix} \cos \alpha & -\sin \alpha \cr
\sin \alpha & \cos \alpha \end{matrix} \right]\left[
\begin{matrix} \eta \cr \phi_1 \end{matrix} \right]\ea with
\begin{align}
\tan 2 \alpha &= {2 v_H m_{\eta \Phi} \over
m_{\phi_1}^2-m_\eta^2}\label{2ALPHA}\\ m^2_{\delta_1} &\simeq
m_\eta^2- {(m_{\eta
\Phi} v_H)^2 \over m_{\phi_1}^2-m_\eta^2}\label{Deltaaa1}\\
m^2_{\delta_2} &\simeq m_{\phi_1}^2+ {(m_{\eta \Phi} v_H)^2 \over
m_{\phi_1}^2-m_\eta^2}\ ,\end{align} where in the last two
equations we have taken $(m_{\eta\Phi}
v_H)^2/(m_{\phi_1}^2-m_\eta^2)^2\ll 1$. In other words, the
mixing, $\alpha$ is small and the interactions of lightest scalar
$\delta_1$ with the  $W$ and $Z$-bosons are suppressed by $\sin
\alpha$ but $\delta_2$ approximately corresponds  to the real
component of the electroweak doublet $\Phi$. Direct searches
\cite{neutralHiggs} restrict $\delta_2$ and $\phi_2$ to be heavier
than $\sim 90$ GeV. On the other hand, $\delta_1$ can be light and
play the role of the SLIM described in the previous section. To
see this more clearly, let us add the couplings  with fermions:
\be \mathcal{L}=-g_{i \alpha}\bar{N}_{i} \Phi^\dagger\cdot
L_\alpha -{M_{i}\over 2} \bar{N}_i^c N_i\ \label{sterileL},\ee
where $L_\alpha$ is the lepton doublet of flavor $\alpha$:
$L_\alpha^T=(\nu_{L\alpha} \ \ell_{L\alpha}^-)$. We focus on the
following range of parameters:
 \be \label{con1} m_{\delta_1}^2 <m_{N_1}^2\ll
m_{\delta_2}^2\simeq m_{\phi_2}^2 \simeq m_{\phi^-}^2 \sim
m_{electroweak}^2  \ee {\rm and} \be \label{con2}\left|
{m_{\phi_2}^2-m_{\delta_2}^2 \over
m_{\phi_2}^2+m_{\delta_2}^2}\right| \simeq \left|-{\lambda_2 \over
2}{v_H^2 \over m_{\phi_2}^2} -{\sin^2 \alpha \over 2}\right| \ll 1
\ . \ee $\delta_1$, which  is called SLIM, is the dark matter
candidate. The main annihilation mode of DM is to neutrino
(antineutrino) pair: \be \langle \sigma(\delta_1\delta_1 \to
\nu_{L\alpha} \nu_{L \beta}) v_r\rangle =\langle
\sigma(\delta_1\delta_1 \to \bar\nu_{L\alpha} \bar\nu_{L \beta})
v_r\rangle= {\sin^4 \alpha \over 8 \pi}\left| \sum_i {g_{i\alpha }
g_{i\beta } m_{N_i} \over m_{\delta_1}^2+m_{N_i}^2} \right|^2 \ .
\ee Considering that $m_{\delta_1}<m_{N_1}$, from this formula we
expect the lightest right-handed neutrino to be  one of the main
contributors to the annihilation cross section so we find
 \be  {\rm Max}[g_{1 \beta}]\sin \alpha
\sim 5 \times 10^{-4}\left( {m_{N_1} \over {\rm MeV}}\right)^{1/2}
\left( {\langle \sigma v_r \rangle \over 3 \cdot 10^{-26} {\rm
cm}^3{\rm sec}^{-1}}\right)^{1/4} (1+{m_{\delta_1}^2 \over
m_{N_1}^2})^{1/2}\ . \label{ggg}\ee Through  a one-loop diagram
active neutrinos acquire the following mass
\begin{align}  (m_\nu)_{\alpha \beta}=& \sum_i
{g_{i\alpha }g_{i\beta } \over 32 \pi}m_{N_i}\Big[ \sin^2 \alpha
({m_{\delta_2}^2 \over m_{N_i}^2-m_{\delta_2}^2}\log {m_{N_i}^2
\over m_{\delta_2}^2} - {m_{\delta_1}^2 \over
m_{N_i}^2-m_{\delta_1}^2}\log {m_{N_i}^2 \over m_{\delta_1}^2})
\nonumber\\
&+{m_{\phi_2}^2 \over m_{N_i}^2-m_{\phi_2}^2}\log {m_{N_i}^2 \over
m_{\phi_2}^2}-{m_{\delta_2}^2 \over m_{N_i}^2-m_{\delta_2}^2}\log
{m_{N_i}^2 \over m_{\delta_2}^2}\Big]\ .\label{MnUGen}
\end{align}
This formula resembles the mass formula in
Eq.~(\ref{flavoredmass}) with the difference that after UV
completion, the UV cutoff has disappeared and instead the masses
of the heavy particles, $m_{\phi_2}$ and $m_{\delta_2}$ show up in
the formulas for the active neutrinos. For
$m_{\delta_1}^2<m_{N_i}^2\ll m_{\delta_2}^2$, we find \be
(m_\nu)_{\alpha \beta}\simeq\sum_i {g_{i\alpha }g_{i\beta } \over
32 \pi}m_{N_i}\Big[ \sin^2 \alpha({m_{\delta_2}^2 \over
m_{N_i}^2-m_{\delta_2}^2}\log {m_{N_i}^2 \over m_{\delta_2}^2} -
{m_{\delta_1}^2 \over m_{N_i}^2-m_{\delta_1}^2}\log {m_{N_i}^2
\over m_{\delta_1}^2}-1)-{\lambda_2}{v_H^2 \over m_{\phi_2}^2}
\Big]\ .\label{mNu}\ee  We can divide the parameter space to the
following two separate regimes : (1) $\lambda_2
v_H^2/(m_{\phi_2}^2) \gg \sin^2\alpha \log(m_{\delta_2}^2/m_N^2)$;
(2) $\lambda_2 v_H^2/( m_{\phi_2}^2) \sim \sin^2 \alpha \log
(m_{\delta_2}^2/m_N^2)$ or $\lambda_2 v_H^2/( m_{\phi_2}^2) \ll
\sin^2\alpha \log(m_{\delta_2}^2/m_N^2)$. In the first case,
Eq.~(\ref{ggg}) combined with Eq.~(\ref{mNu}) implies
$m_{N_1}\ll1$ which is disfavored by big bang nucleosynthesis
\cite{serpicoraffelt}. For  $\lambda_2 v_H^2/(
m_{\phi_2}^2)\stackrel{<}{\sim} \sin^2 \alpha \log
(m_{\delta_2}^2/m_N^2)$, we find $$m_{\delta_1} \ll m_{N_1}\sim
{\rm few~MeV}$$ which is the same condition as we found for the
low energy SLIM scenario in the previous section.

Notice that within this model no upper bound on the masses of
$\phi^-$, $\delta_2$ or $\phi_2$ is found. However, as we will see
below, for relatively  light $\Phi$, the model is more natural.
From Eq.~(\ref{ggg}) and the perturbativity of $g_{1 \beta}$, we
find that $\sin \alpha$ cannot be smaller than $\sim 10^{-4}$. For
$\sin \alpha\ll 1$ and $m_{\Phi}^2 \gg m_\eta^2$, we can write
(see Eqs. (\ref{2ALPHA},\ref{Deltaaa1}) \be \label{app}
m_{\delta_1}^2 \simeq m_\eta^2-\sin^2\alpha m_\Phi^2\ .\ee Thus,
for $m_\Phi \gg 100$~GeV, a fine tuned cancelation between the two
terms in Eq.~(\ref{app}) is required to maintain $m_{\delta_1}$
below 10 MeV. In other words, based on naturalness of the model we
expect $\phi^-$, $\delta_2$ or $\phi_2$ to be within the reach of
the LHC. We shall discuss this point in section \ref{LHC}.

Notice that this model has some features in common with the
so-called inert model \cite{inert} but  with the difference that
here we have an extra singlet scalar and the main annihilation
mode of dark matter pair is into neutrinos. The contribution to
the oblique parameters in our model is however similar to that in
the inert model. Similarly to the inert model, within this model
the SM Higgs can be as heavy as a few 100 GeV without violating
bounds from the electroweak precision data. That is  because the
contribution from the new doublet to the oblique parameters can
cancel the one from a SM Higgs.

\section{Implications of the SLIM model\label{imp}}

The impact of this model on the low energy phenomena such as the
decay of light mesons and supernovae is similar to the low energy
scenario discussed in section \ref{SLimplication} except that the
coupling $g_{i \beta}$ has to be replaced by $g_{i \beta} \sin
\alpha /\sqrt{2}$. However, since this model also contains new
heavy states, its phenomenology is richer. In particular, the
heavy states can be produced at the LHC. We will discuss about
this in more detail in section \ref{LHC}. Here, we discuss the
impact that this model can have on other phenomena: (1)
Annihilation modes of DM to an electron positron pair or photon
pair; (2) Dark matter self-interaction;  (3) Magnetic dipole
moment of the muon; (4) LFV rare decay of charged lepton. As we
shall see, bounds on rare decay already constrains a part of the
parameter space. Let us discuss them one by one.
\subsection{Annihilation into electron positron pair and photon
pair}
The annihilation to the $e^- e^+$ pair is loop suppressed by a
factor of $e^4 /(16 \pi^2 \sin^2\theta_W)^4$ \cite{myModel}.
Because of this suppression, the rate of ${\rm DM+DM}\to e^-e^+$
is too low to account for the disputed 511 keV signal. Moreover
the flux of radiation from the  $e^- e^+$ pair would be too low to
be detectable. Of course, in this model the DM is too light to
annihilate to $\mu^- \mu^+$ pair.

At one-loop level a pair of $\delta_1$ can also annihilate into a
photon pair with cross section \be \sigma(\delta_1\delta_1 \to
\gamma \gamma)\sim{e^8 \sin^4 \alpha \over 8 \pi (16
\pi^2)^2\cos^4 \theta_W}{m_{\delta_1}^2 \over m_W^4}\sim{\rm few}
\times 10^{-41} \left( {M_{\delta_1}\over {\rm MeV}}\right)^2
\sin^4 \alpha~{\rm cm}^3/{\rm sec} \ . \ee Because of the loop
suppression, the flux of photons would be too small to be
detectable at Fermi telescope  (see, {\it e.g.,} Fig 4 of
\cite{stefano}).
\subsection{Self-interaction of Dark Matter}
The $\lambda^\prime$ couplings in Eq.~(\ref{MainLag}) can lead to
the self-interaction of the DM pairs with the following cross
section
$$ \langle \sigma(\delta_1\delta_1\to \delta_1\delta_1) v\rangle
\sim {\rm Max}[{|\lambda'_1|^2 \sin^4 \alpha\over 8 \pi
m_{\delta_1}^2},{|\lambda'_2|^2 \cos^4 \alpha\over 8 \pi
m_{\delta_1}^2},{|\lambda'_3|^2 \sin^2\alpha \cos^2 \alpha \over 8
\pi m_{\delta_1}^2}]\ .$$ The self-interaction of DM
 is constrained by merging galaxy clusters \cite{bullet}:
 $\sigma/m_{DM}\stackrel{<}{\sim}1~{\rm cm^2/g}$ which translates
 into
 $$|\lambda_1'|^2 \sin^4\alpha,|\lambda'_2|^2\cos^4 \alpha,|\lambda'_3|^2
\sin^2\alpha \cos^2 \alpha\stackrel{<}{\sim} 10^{-4}\ .$$

\subsection{$(g-2)_\mu$ \label{MMDM}}
Via coupling  $g_{i \mu } \bar{N}_i \mu_L (\phi^-)^\dagger$, muons
receive a magnetic dipole moment at one-loop level:
$$ \delta{g-2\over 2}=\sum_i {|g_{i\mu }|^2 \over 16
\pi^2}{m_\mu^2 \over m_{\phi^-}^2}K(t_i)\ ,$$ where
 \be K(t_i)={2 t_i^2+5t_i-1 \over
12 (t_i-1)^3}-{t_i^2 \log t_i \over 2(t_i-1)^4} \ ,\label{k(t)}
\ee in which $t_i=m_{N_i}^2/m_{\phi^-}^2.$ For $t_i \ll 1$,
$$\delta{g-2\over 2}=5\times 10^{-12}{\sum_i |g_{i \mu }|^2 \over
10^{-2}}\left({100~{\rm GeV} \over m_{\phi^-}^2}\right)^2$$ which
is two orders of magnitude below the present bound.

\subsection{ LFV rare decay}
The coupling $g_{i \alpha} \bar{N}_i \ell_{L\alpha}\phi^-$ leads
to
 the Lepton Flavor Violating rare decays, $\mu \to e\gamma$,
$\tau \to \mu \gamma$ and $\tau \to e \gamma$. From the formulas
in \cite{Lavoura}, we find\be \Gamma(\ell_\alpha \to \ell_\beta
\gamma)={m_\alpha^3 \over 16\pi}|\sigma_R|^2\ , \ee where
$$\sigma_R=\sum_i g_{i \alpha } g_{i\beta }^* {iem_\alpha \over 16
\pi^2 m_{\phi^-}^2} K(t_i)\ , $$ where
$t_i=m_{N_i}^2/m_{\phi^-}^2$ and $K(t_i)$ is defined in
Eq.~(\ref{k(t)}). Within this model $t_1\ll 1$ however $t_2$ can
be either small or larger than 1. In case that $t_2$ is also small
we can write
\begin{align}
{\rm Br}(\mu \to e \gamma)&\sim 2 \times 10^{-4}|\sum_i g_{\mu
i}g_{e i}^*|^2 \left({100~{\rm GeV} \over m_{\phi^-}}\right)^4 \\
{\rm Br}(\tau \to \ell_\alpha \gamma)&\sim 5 \times 10^{-5}|\sum_i
g_{i\tau }g_{i\alpha }^*|^2 \left({100~{\rm GeV} \over
m_{\phi^-}}\right)^4.
\end{align}
The latest bound \cite{pdg} on these branching ratios are
\begin{align}
{\rm Br}(\mu \to e \gamma)&<1.2\times 10^{-11} \\
{\rm Br}(\tau \to e \gamma)&< 3.3 \times 10^{-8} \\
{\rm Br}(\tau \to \mu \gamma)&<4.4 \times 10^{-8}\ .
\end{align}
These bounds already excludes large values of the couplings.
However, the following range is consistent with bounds and is
particularly interesting from phenomenological point of view as in
this range the forthcoming LFV searches have a good chance of
observing a signal:
 \be m_{\phi^-}\sim 100~{\rm GeV} \ \ \
g_{i \mu},g_{i\tau}\sim{\rm few}\times 10^{-2} \  \ {\rm and} \ \
g_{i e}\sim{\rm few}\times 10^{-3}.\label{gParticular}\ee
\section{Signature of the SLIM model at the LHC\label{LHC}}

As we discussed in the previous section, naturalness implies that
the new particles $\phi^-$, $\delta_2$ and $\phi_2$ have masses
not much higher than $O(100~{\rm GeV})$. As a result, these
particles are expected to be produced in pairs at the LHC. The
produced $\phi^-$ can decay via its coupling in
Eq.~(\ref{sterileL}) to  charged leptons:
 \be \Gamma(\phi^- \to l_\beta N_i)  =\frac{|g_{i
\beta}|^2}{ 16 \pi}
\frac{(m_{\phi^-}^2-m_{N_i}^2)^2}{m_{\phi^-}^3}~~~~~{\rm for} ~~
m_{N_i}<m_{\phi^-}-m_{l_\beta} \ . \ee Notice that the same
couplings that determine the flavor structure of the neutrino mass
matrix as well as branching ratios of the LFV rare decays of the
charged leptons also determine these decays. This fact provides a
means for the cross check of the model. In \cite{Hashemi}, the
feasibility of determining the coupling at the LHC has been
studied considering the various sources of background and
employing state-of-the-art techniques to enhance the signal to
background ratio. In this section, we review the results of this
analysis. The details of the analysis and the software that has
been used to perform this analysis can be found in \cite{Hashemi}.

As we saw in section \ref{UV}, at least the lightest $N_i$, which
we call $N_1$ has to be light which means the decay modes  $\phi^-
\to N_1 e^-$, $N_1\mu^-$ and $N_1\tau^-$  are all kinematically
allowed. However, the second right-handed neutrino can be heavier.
The following three situations are possible: \begin{itemize} \item
$m_{N_2}>m_{\phi^-}$. In this case, only $\phi^- \to N_1
l_\beta^-$ will be possible.
\item $m_{N_2}<m_{\phi^-}$ and $m_{N_2}\sim m_{\phi^-}$. In this
case, both $\phi^- \to N_1 l_\beta^-$ and $\phi^- \to N_2
l_\beta^-$ are kinematically possible and by studying the energy
spectrum of the charged lepton, these two decay modes can  in
principle be distinguished. However, considering that the
contribution of $N_2$ to the neutrino mass cannot be larger than
about $\sim \sqrt{\Delta m_{atm}^2/\Delta m_{sun}^2}\sim 10$ times
the contribution of $N_1$, Eq.~(\ref{MnUGen}) implies that
$${\rm for}~ m_{N_2}\sim m_{\phi^-}, \ \ \ \ \ g_{2\beta}\ll g_{1
\beta}  \ .$$ This means for this situation,  $\phi^- \to N_1
l_\beta^-$ will dominate over $\phi^- \to N_2 l_\beta^-$. As a
result, the signal for $\phi^- \to l_\beta^-+{\rm missing~energy}$
will be mostly composed of $\phi^- \to N_1 l_\beta^-$.
\item $m_{N_2}\ll m_{\phi^-}$. In this case, the masses of both $N_1$ and $N_2$
can be neglected and the energy of $l_\beta$ in both cases will be
approximately equal to $m_\phi^-/2$ in the rest frame of $\phi^-$.
Thus, the signals for $\phi^- \to N_1 l_\beta^-$ and $\phi^- \to
N_2 l_\beta^-$ cannot be distinguished. The signal for $\phi^- \to
l_\beta^-+{\rm missing~energy}$ is determined by the following sum
$$|g_{1 \beta}|^2+|g_{2 \beta}|^2.$$
\end{itemize}
In \cite{Hashemi} only the case $m_{N_2}>m_{\phi^-}$ is studied
but from the above discussion, we conclude that the analysis in
\cite{Hashemi} also applies for the case $m_{N_1}\ll
m_{N_2}<m_{\phi^-}$ because $\phi^-$ practically only decays to
$N_1$ just like the case $m_{N_2}>m_{\phi^-}$.  For the case
$m_{N_1}\sim m_{N_2}$, the analysis of \cite{Hashemi} also applies
but one has to replace $|g_{1 \beta}|^2$ with $|g_{1
\beta}|^2+|g_{2 \beta}|^2.$

To make the analysis simpler it has been assumed in \cite{Hashemi}
that $m_{\phi_2}-m_{\phi^-}$, $m_{\delta_2}-m_{\phi^-}$ and
$|m_{\delta_2}-m_{\phi_2}|$ do not exceed 80 GeV to forbid two
body decays $\delta_2,\phi_2 \to W^+\phi^-$ or $\phi_2 \to
\delta_2 Z $ (or $\delta_2 \to \phi_2 Z $). Moreover, $\phi^-$ can
have other decay modes such as $\phi^- \to W^- \delta_1$,
$\phi^-\to \delta_1 \ell_\alpha^- \nu$ and $\phi^-\to \delta_1 W^-
\gamma$ but these modes can be neglected for $g_{1
\alpha}\stackrel{>}{\sim} 0.01$ and $\sin \alpha\stackrel{<}{\sim}
0.01$.

\begin{table}
\begin{center}
\begin{tabular}{|l|c|c|}
\hline
& Point A & Point B \\
\hline
$m_{N_{1}}$~({\rm MeV}) & 1 & 1 \\
\hline
$m_{N_{2}}$ & $> m_{\phi^-}$ & $>m_{\phi^-}$ \\
\hline
$\alpha$ & 0.01 & 0.01 \\
\hline
$\lambda_{2}$ & 0 & 0 \\
%
$m_{\phi_2}$~({\rm GeV}) & 90 & 90 \\
\hline
$\theta$ & $\pi / 2$ & 0 \\
\hline\hline
$g_{1\alpha }$ & $\left( \begin{array}{c} 0 \\ 0.03 \\ 0.03 \end{array} \right)$ & $\left( \begin{array}{c} 0.01 \\ 0.01 \\ -0.01 \end{array} \right)$ \\
\hline
\end{tabular}
\end{center}
\caption{Model parameters.\label{par}}
\end{table}
\cite{Hashemi} studies the pair production of $\phi^+\phi^-$,
$\phi^\pm\phi_2$ and $\phi^\pm\delta_2$ at the LHC and the
subsequent decay of $\phi^\pm$ to charged leptons. To perform the
analysis, the two benchmark points with parameters shown in table
\ref{par} have been studied. At point A,
$\textnormal{BR}(\phi^{\pm}\rightarrow e^\pm N_1)=0$ and
$\textnormal{BR}(\phi^{\pm}\rightarrow \mu^\pm
N_1)\simeq\textnormal{BR}(\phi^{\pm}\rightarrow \tau^\pm N_1)
\simeq 0.5$. At point B, $\textnormal{BR}(\phi^{\pm}\rightarrow
e^\pm N_1)\simeq \textnormal{BR}(\phi^{\pm}\rightarrow \mu^\pm
N_1)\simeq\textnormal{BR}(\phi^{\pm}\rightarrow \tau^\pm N_1)
\simeq 1/3$. The main focus in \cite{Hashemi} is on point A.

The cross section of the $\phi^+\phi^-$ production at the 14 TeV
run of the LHC for $m_{\phi^\pm}$ between 80~GeV to 130~GeV varies
between 800 fb to 200 fb. At the benchmark point A, the subsequent
decay of $\phi^\pm$ will lead to four types of signal
$\mu^+\mu^-+{\rm missing~ energy}$, $\tau^+\mu^-+{\rm missing~
energy}$, $\mu^+\tau^-+{\rm missing~ energy}$ and
$\tau^+\tau^-+{\rm missing~ energy}$. The $\tau \tau$ final state
is contaminated by large hadronic backgrounds such as $W+{\rm
jets}$ followed by the  decay of $W$ to jets or $\tau$. Light jets
fake $\tau$ even after applying the cuts so this mode has not been
discussed in \cite{Hashemi}.  In \cite{Hashemi}, the sum of
signals $\tau^+\mu^-+{\rm missing~ energy}$ and $\mu^+\tau^-+{\rm
missing~ energy}$ has been collectively studied.

The main sources of background are $W^{+}W^{-}$, $t\bar{t}$,
$W+$jets and $Z+$jets. Their cross sections at 14 TeV center of
mass energy are shown in Table \ref{bXsec}. Like the
$\phi^+\phi^-$ signal, the $W^+W^-$ pair can lead to $l^+_\beta
l^-_\gamma+{\rm missing~energy}$. Notice that the cross section of
the $W^+W^-$ production is about two orders of magnitude higher
than that of the $\phi^+\phi^-$ production. Moreover,
misidentification of some of the jets or other misidentifications
can lead to mimicking the signal. To reduce the background and
therefore enhance the signal significance ({\it i.e.,} ${\rm
signal}/\sqrt{\rm background}$), several cuts are suggested in
\cite{Hashemi}. Using these cuts, explicit computation of the
signal significance is carried out for benchmark point A  at 14
TeV run of the LHC and for 30 fb$^{-1}$. The results are displayed
for the $\tau \mu+{\rm missing~ energy}$ and $\mu \mu+{\rm
missing~ energy}$ signals respectively in tables \ref{sig} and
\ref{sig2}. As seen from table \ref{sig2}, for the values of
$m_{\phi^-}$ lower than 130~GeV, the discovery can be made by 30
fb$^{-1}$ of data. Considering that the background is almost the
same,  the signal significance of the $e^-e^++{\rm
missing~energy}$ signal can be obtained by scaling that of $\mu^-
\mu^++{\rm missing~energy}$ by a factor of $${ {\rm Br}(\phi^+ \to
N_1 e^+){\rm Br}(\phi^- \to N_1 e^-)|_{{\rm At~ Point~ B}} \over
{\rm Br}(\phi^+ \to N_1 \mu^+){\rm Br}(\phi^- \to N_1
\mu^-)|_{{\rm At~ Point~ A}}}\simeq {4\over 9} \ .$$ That is at
$m_{\phi^\pm}=80$~GeV and the benchmark point B, the signal
significance of $e^+e^-+{\rm missing~energy}$ for 30 fb$^{-1}$ of
data will be 4.1$\sigma$ C.L. As indicated in \cite{Hashemi}, this
is  a simplified estimation as there might be some difference
between muon and electron reconstruction and selection
efficiencies in the detector.  A detailed study of these features
needs a full simulation of the detector.

A crucial question is whether $g_{i \beta}$ can be derived from
the data. As discussed before, deriving the flavor structure of
$g_{i \beta}$ helps us to cross-check the model as the same
couplings determine the neutrino mass matrix and LFV branching
ratios. To derive $g_{i \beta}$, one should extract the signal
number, $N_S$ which is generally given by
\begin{equation}
 N_{S}=\frac{N_{obs.}-N_{B}}{\epsilon_S}
\end{equation}
where $\epsilon_S$ is  the selection efficiency of the signal,
$N_{obs.}$ is the observed number of events and $N_{B}$ is the
contamination due to the background. $N_B$ is calculated by
simulation using input such as Parton Distribution Functions (PDF)
or total luminosity.   The uncertainty in these inputs induce an
uncertainty  in extracting $N_S$ and therefore the couplings. The
main source of uncertainty is the uncertainty in PDFs which at
present is about 10 \% uncertainty. These uncertainties induce an
uncertainty of about 60 \% in extracting $N_S$. In \cite{Hashemi},
it was shown that $\Delta N_S/N_S$ due to these uncertainties does
not improve by increasing the luminosity. However, increasing the
 center of mass energy will enhance signal to background ratio and
therefore improve $\Delta N_S/N_S$. Of course, if by using the
data of LHC or some other machine, the uncertainty in PDFs are
reduced, the precision of extracting $g_{i \alpha}$ can be
improved.

Other modes that have been discussed in \cite{Hashemi} are $pp\to
\phi^\pm \phi_2$ and $pp\to \phi^\pm \delta_2$. As discussed
earlier in this section, because of the simplifying assumptions on
the mass spectrum of the components of $\Phi$, decay modes such as
$\phi^- \to W^- \delta_1$ are forbidden so $\delta_2$ and $\phi_2$
can have only invisible decay modes $\delta_2,\phi_2 \to N_1 \nu$.
As a result, the signal will be composed of  a charged lepton from
the decay of $\phi^\pm$ plus missing energy which is composed of
$N_1$ from $\phi^\pm$ decay and the decay products of $\phi_2$ or
$\delta_2$. At benchmark point A and for $m_{\phi^\pm}=80$~GeV,
the significance of the $\mu+{\rm missing~energy}$ and  $\tau+{\rm
missing~energy}$ signals can reach as high as $9\sigma$ and
$4.6\sigma$ at 14 TeV energy and 30 fb$^{-1}$ integrated
luminosity \cite{Hashemi}. Again under simplifying assumption that
the selection efficiency of detecting muon and electron is not
much different the significance of the $e+{\rm missing~energy}$
signal at the B point is equal to that of $\mu+{\rm
missing~energy}$ signal at the A point rescaled by a factor of
${\rm Br}(\phi^+ \to N_1 e^+)|_{\rm At~point~B}/{\rm Br}(\phi^+
\to N_1 \mu^+)|_{\rm At~point~A}\simeq 2/3$.

The above results are for the 14 TeV run of the LHC. In
\cite{Hashemi}, an estimation of the signal significance for the 7
TeV is made by rescaling the cross sections of both background and
signal to their values at 7 TeV run.  The results for 30 fb$^{-1}$
are displayed at table \ref{7TeVsig}.

\begin{table}
\begin{center}
\begin{tabular}{|l|c|c|c|c|}
\hline
Process & $W^{+}W^{-}$ & $t\bar{t}$ & W+jets & Z+jets \\
\hline
Cross Section & 115.5$\pm$0.4 pb& 878.7$\pm$0.5 pb& 187.1$\pm$0.1 nb & 258.9$\pm$0.7 nb \\
\hline
\end{tabular}
\end{center}
\caption{Background cross sections \label{bXsec}}
\end{table}

\begin{table}
\begin{center}
\begin{tabular}{|l|c|c|c|c|}
\hline
$m_{(\phi^{\pm})}$ & 80 GeV & 90 GeV & 110 GeV & 130 GeV \\
\hline
Signal significance & 2.8 & 2.2 & 1.4 & 1 \\
\hline
\end{tabular}
\end{center}
\caption{Signal significance in the $\tau\mu E^{miss}_{T}$ final
state for different $m_{(\phi^{\pm})}$ hypotheses at
$30~fb^{-1}$.\label{sig}}
\end{table}
\begin{table}
\begin{center}
\begin{tabular}{|l|c|c|c|c|}
\hline
$m_{(\phi^{\pm})}$ & 80 GeV & 90 GeV & 110 GeV & 130 GeV \\
\hline
Signal significance & 9.2 & 8.4 & 6.6 & 4.9 \\
\hline
\end{tabular}
\end{center}
\caption{Signal significance in $\mu\mu E^{miss}_{T}$ final state
for different $m_{(\phi^{\pm})}$ hypotheses.\label{sig2}}
\end{table}

\begin{table}
\begin{center}
\begin{tabular}{|l|c|c|}
\hline
Channel & Mass Point & Signal significance \\
\hline \multirow{4}{*}{$\phi^{+}\phi^{-}\rightarrow \tau\mu
E^{miss}_{T}$ } &
$m_{(\phi^{\pm})}=80$ GeV & 1.6\\
& $m_{(\phi^{\pm})}=90$ GeV & 1.2 \\
& $m_{(\phi^{\pm})}=110$ GeV & 0.7 \\
& $m_{(\phi^{\pm})}=130$ GeV & 0.5 \\
\hline \multirow{4}{*}{$\phi^{+}\phi^{-}\rightarrow \mu\mu
E^{miss}_{T}$ } & $m_{(\phi^{\pm})}=80
$ GeV & 6.4 \\
& $m_{(\phi^{\pm})}=90$ GeV & 5.7 \\
& $m_{(\phi^{\pm})}=110$ GeV & 4.2 \\
& $m_{(\phi^{\pm})}=130$ GeV & 3.0 \\
\hline
$\phi^{\pm}\phi_{2}\rightarrow \tau E^{miss}_{T}$ & $m_{(\phi^{\pm})}=80$ GeV & 2.6 \\
\hline
$\phi^{\pm}\phi_{2}\rightarrow \mu E^{miss}_{T}$ & $m_{(\phi^{\pm})}=80$ GeV & 5.0 \\
\hline
\end{tabular}
\end{center}
\caption{Signal significance in different final states for the 7
TeV run, provided that 30 $fb^{-1}$ data is collected at this
energy before any switch to higher machine
energies.\label{7TeVsig}}
\end{table}
\section{AMEND\label{mend}}
In this section, we review the AMEND model which was introduced in
\cite{AMEND}. AMEND stands for A Model Explaining Neutrino masses
and Dark matter. Like the SLIM model, there is a $Z_2$ symmetry
that protects DM from decay. The $Z_2$ symmetry also forbids Dirac
mass at one-loop level. The $Z_2$ symmetry in this model is the
remnant of a global U(1)$_X$ symmetry which is explicitly broken
by small parameters. In the limit of exact $U(1)_X$, neutrino
masses vanish. Neutrino masses are suppressed both by a loop
factor and the small $U(1)_X$ breaking term ({\it i.e.,} `t Hooft
criterion). The particle content of the model includes two
fermionic doublets $R$ and $R^\prime$ with opposite hypercharges,
an electroweak triplet $\Delta$ and a complex singlet, $\phi$.
These particles can in principle be produced at colliders. In
particular, one of the components of $\Delta$ is doubly charged
and can lead to almost background free signal of same sign charged
lepton pair plus missing energy via $\Delta^{++} \to l^+_\alpha
 l^+_\beta \delta_{1,2}$.
The LFV couplings of these particles lead to LFV rare decays such
as $\mu \to e \gamma$, $\tau \to e \gamma$ and  $\tau \to
\mu\gamma$ at loop level.

Within this model, the DM production in the early universe is
thermal. In \cite{AMEND}, various possible annihilation modes have
been discussed. It was found that the dominant annihilation mode
was the DM annihilation through $s$-channel Higgs exchange which
can account for the observed density of DM. DM in this model can
be counted as a Weakly Interacting Massive Particle (WIMP) and can
 show up in direct DM search experiment based on measuring recoil
energy from scattering DM off nuclei in a background free
environment. Various experiments are designed for this purpose.
Their results are contradictory. On one hand, experiments such as
XENON do not find any signal and on the other hand, the DAMA
experiment \cite{Dama} reports a positive signal at more than 8
$\sigma$ C.L. To reconcile these conflicting results, several
attempts have been made. Among these solutions, inelastic DM
solution \cite{TuckerSmith} and light DM ($<10$~ GeV)
\cite{lightDM} have received more attention. In \cite{AMEND}, null
results interpreted as an upper bound on the cross-section as well
as the two solutions accommodating the positive signal from direct
DM searches have been studied and shown that they can be embedded
within the AMEND model by going to the proper regions of the
parameter space. However, more recent data from XENON100
\cite{XENON100} and a re-analysis of XENON10 data \cite{XENON10}
respectively disfavor the inelastic DM and light DM solutions.  We
therefore focus on the constraint from DM searches.

In this section, we  first describe AMEND and then discuss its
implications for various observations.

\subsection{Description of AMEND \label{AMENDmodel}}
\begin{table}
\begin{center}
\begin{tabular}{|l|ccc|l|}\hline
particle & ${SU}({3})_c$ & ${SU}({2})_L$ & ${U}({1})_Y$&
 \\\hline\hline
$Q_L$ & 3 & 2 & 1/6  &\multirow{7}{*}{fermion}\\
$u_R$ & 3 & 1 & 2/3  & \\
$d_R$ & 3 & 1 & -1/3 &  \\
$\ell_L$ & 1 & 2 & -1/2 &\\
$e_R$ & 1 & 1 & -1  & \\
$R=R_R$ & 1 & 2 & -1/2 &\\
$R^\prime=R^\prime_R$ & 1 & 2 & 1/2&
 \\\hline
$H$ & 1 & 2 & 1/2    &\multirow{3}{*}{scalar} \\
$\Delta$ & 1 & 3 & 1 &\\
$\phi$ & 1 & 1 & 0  &\\\hline
\end{tabular}
\caption{Particle content and gauge quantum
numbers\label{tab:partcont}.}
\end{center}
\end{table}

The particle content of this model is listed in table
\ref{tab:partcont}. As seen from this table, the new particles are
the following.
\begin{itemize}
\item A complex scalar field which can be decomposed  in terms of real fields
as $\phi\equiv (\phi_1+i\phi_2)/ \sqrt{2}$; \item A scalar triplet
with the following components
\begin{equation}
 \Delta=\left[
\begin{matrix}
\frac{\Delta^+}{\sqrt{2}} & \Delta^{++} \cr \Delta^0
&-\frac{\Delta^+}{\sqrt{2}}
\end{matrix}\right]\ . \label{eq:components}
\end{equation}
We can write $\Delta^0=(\Delta_1+i\Delta_2 )/\sqrt{2} $ where
$\Delta_1$ and $\Delta_2$ are real scalar fields.

\item Two Weyl fermion $SU{2}_L$ doublets, $R^T=(\nu_R \ E_R^-)$ and $(R^\prime)^T=(E_R^+ \
\nu_R^\prime)$.
\end{itemize}
A symmetry, called $G$ symmetry, is defined under which each of
the new particles are charged under a separate U(1). The $G$
symmetry is defined as follows.
\begin{equation} \label{eq:Group}
 G \equiv {\rm U}(1)_R \times {\rm U}
 (1)_\phi \times {\rm U}(1)_\Delta
 \times
 {\rm U}(1)_\ell\;
\end{equation}
where  ${\rm U}(1)_\phi$ and ${\rm U}(1)_\Delta$  are symmetries
under which only $\phi$ and $\Delta$ are respectively charged and
${\rm U}(1)_R$ is the symmetry under which $R$ and $R^\prime$ have
opposite quantum numbers. ${\rm U}(1)_\ell$ is the familiar lepton
number ${\rm U}(1)$ symmetry associated with lepton number. The
model is constructed such that the  main part of its Lagrangian
preserves the $G$ symmetry.
  In addition to the kinetic and the  gauge interaction terms,
  the most general $G$-preserving Lagrangian is composed of  the
  following scalar potential

 \begin{equation} \label{eq:Vs}
\begin{split}
\mathcal{V}=&-\mu_H^2 H^\dagger H + \mu_\Delta^2\
Tr\left(\Delta^\dagger
 \Delta\right) + \mu_\phi^2 \phi^\dagger \phi\\
 &+\frac{\lambda}{4} (H^\dagger H)^2 + \frac{\lambda_\phi}{4}
 (\phi^\dagger \phi)^2
 +\frac{\lambda_{\Delta1}}{2}\left(\ Tr\Delta^\dagger\Delta\right)^2+
 \frac{\lambda_{\Delta2}}{2}\ Tr (\Delta^\dagger [ \Delta^\dagger, \Delta ]
 \Delta)\\
& + \lambda_{H\Delta1}H^\dagger H \
Tr\left(\Delta^\dagger\Delta\right)+
 \lambda_{H\Delta2}H^\dagger[\Delta^\dagger\ ,\Delta] H + \lambda_{\phi\Delta}
 \phi^\dagger\phi \ Tr \left(\Delta^\dagger \Delta\right) + \lambda_{H\phi} \phi^\dagger\phi
 H^\dagger H ~,
\end{split}
\end{equation}
and the fermionic part which is a Dirac mass term for the
fermionic doublet
\begin{equation}
-\mathcal{L}_R = m_{RR} (R^{\prime C})^\dagger \cdot R+\hc   ~,
\label{eq:bb}
\end{equation}
where $(R^{\prime C})^T=(\nu_R^{\prime C}\ -(E_R^+)^C )$. $m_{RR}$
should be heavier than $\sim 100$ GeV to avoid the bounds from
direct searches. As a reference point, we shall take $m_{RR}=300$
GeV. At a very high energy scale, the $G$ symmetry breaks to a
smaller U(1)$_X$ symmetry under which the SM particles are all
neutral and the quantum numbers of the new particles are as
follows.
$$ R\stackrel{U(1)_X}{\Longrightarrow} +1 \  , \ R^\prime\stackrel{U(1)_X}{\Longrightarrow}
-1\ , \ \Delta \stackrel{U(1)_X}{\Longrightarrow} +1 \ {\rm and} \
\phi \stackrel{U(1)_X}{\Longrightarrow} -1.$$ Notice that U(1)$_X$
is anomaly-free and can in principle be fixed. The terms that
break $G$ to its $U(1)_X$ subgroup are the following
\begin{subequations} \label{eq:U1XL}
\begin{align}
\mathcal{V}_{H\Delta\phi}=& \lambda_{H\Delta\phi} H^T\I\sigma_2
\Delta^\dagger H \phi^\dagger +\hc \\
\mathcal{V}_{\ell_L \phi} = & g_\alpha \phi^\dagger
R^\dagger\ell_{L\alpha} +\hc ~. \label{eq:fmess}
\end{align}
\end{subequations}
At a lower energy scale,  U(1)$_X$ breaks to a $Z_2$ symmetry
under which the SM particles are even but the new particles are
all odd. After $U(1)_X \to Z_2$, the Lagrangian includes the
following terms for the scalars
\begin{equation} \label{eq:tildeVs}
\widetilde{\mathcal{V}}_{{\rm scalar}}=\tilde
{\lambda}_{H\Delta\phi} H^T \I\sigma_2 \Delta^\dagger H \phi+
\tilde {\mu}_\phi^2 \phi^2 + \tilde {\lambda}_{\phi\,1} \phi^4 +
\tilde {\lambda}_{\phi\,2} \phi^3 \phi^\dagger + \tilde
{\lambda}_{H\phi} H^\dagger H \phi^2 + \tilde
{\lambda}_{\Delta\phi} \tr \Delta^\dagger \Delta \phi^2 +\hc\; .
\end{equation}
and the following terms for the fermions
 \begin{equation}
\label{eq:fmessvio} -\widetilde{\mathcal{L}}_{\ell_L \phi }=
\tilde{g}_\alpha \phi R^{\dagger} \ell_{L\alpha} +\hc  \ \ {\rm
and } \ \ - \widetilde{\mathcal{L}}_{\ell_L \Delta} =
(\tilde{g}_\Delta)_\alpha R^{\prime\dagger} \cdot \Delta \cdot
\ell_{L\alpha}+\hc \
\end{equation} The pattern of symmetry breaking implies
 \mbox{$g \gg \tilde{g}, \tilde{g}_\Delta$} and \mbox{${\lambda}_{H\Delta\phi}
\gg \tilde{\lambda}_{H\Delta\phi}$.}

\subsection{The scalar sector \label{ssector}}
Within this model, only the SM Higgs obtains a vacuum expectation
value and
\begin{equation}\label{eq:VEVconf}
\langle \phi_1\rangle=\langle \phi_2 \rangle =\langle
\Delta_1\rangle=\langle \Delta_2 \rangle=0\ .
\end{equation}

 After electroweak symmetry breaking, $\phi$ and
$\Delta^0$ mix with each other. For the CP-symmetric case, CP even
scalars $\phi_1$ and $\Delta_1$ mix only with each other and the
CP-odd scalars $\phi_2$ and $\Delta_2$ mix among each other. The
neutral scalar mass eigenstates, $\delta_1$,  $\delta_2$,
$\delta_3$ and $\delta_4$ can be written in terms of the
components of $\phi$ and $\Delta^0$ as follows
\begin{equation} \label{eq:deltaS}
\left(\begin{array}{c} \delta_1\\\delta_2\\\delta_3\\\delta_4
\end{array}\right)=\left(\begin{array}{cccc}
\cos\alpha_1 & 0 & \sin\alpha_1 & 0\\
0 & \cos\alpha_2 & 0 & \sin\alpha_2\\
-\sin\alpha_1 & 0 & \cos\alpha_1 & 0\\
0 & - \sin\alpha_2 & 0 & \cos\alpha_2\\
\end{array}\right)
\left(\begin{array}{c} \phi_1 \\ \phi_2 \\ \Delta_1 \\ \Delta_2
\end{array}
\right) ~,
\end{equation}
 where $|\tan 2 \alpha_1| \simeq |\tan 2 \alpha_2 |
 \simeq {2 m_{\phi \Delta}^2}/({m_\Delta^2 - m_\phi^2})$. In the following the
masses of $\delta_i$ are denoted by $M_i$. The formula for $M_i$
can be found in \cite{AMEND}.
 The difference $\left||\alpha_1|-|\alpha_2|\right|$ as well as the mass
splittings $|M_2-M_1|$ and $|M_4-M_3|$
  are suppressed  by the $\U{1}_{X}$-breaking terms.
We take $\delta_1$ to be the lightest new particle and therefore a
DM candidate. As discussed in \cite{AMEND}, the CP-odd scalar
$\delta_2$ could also play the role of DM.

The coupling of $\delta_1$ to the $Z$-boson is of form
\begin{equation}
 \label{eq:z-coupling} \frac{\I\, g_{\SU{2}} \sin \alpha_1\sin
\alpha_2}{ \cos \theta_W} [\delta_2
\partial_\mu \delta_1-\delta_1 \partial_\mu \delta_2]Z^\mu  ~,
\end{equation}
where $g_{\SU{2}}$ is the SM weak gauge coupling and $\theta_W$ is
the Weinberg angle. If $M_1+M_2<m_Z$, \be \Gamma(Z\to
\delta_1\delta_2)=\frac{G_F
  \sin^2\alpha_1\sin^2\alpha_2}{6\sqrt{2}\pi}m_Z^3 ~.
\ee $\delta_2$ will eventually decay to $\delta_1$ and a neutrino
pair so this decay mode will count as an extra contribution to
invisible decay mode of the $Z$ boson.
 For $M_1+M_2<m_Z$, the upper bound on the  extra invisible decay
 modes of the $Z$ boson \cite{PDG} implies $$\sin \alpha_1 \sin
 \alpha_2<0.07$$

\subsection{Neutrino masses and LFV rare decays\label{NUmass}}
Within this model,  there  are one-loop contributions to the
neutrino mass matrix of form \cite{AMEND}
\begin{equation} \label{mmmnnnn}
(m_\nu)_{\alpha \beta} = [g_\alpha
(\tilde{g}_\Delta)_\beta+g_\beta (\tilde{g}_\Delta)_\alpha]
\tilde{\eta} + [\tilde{g}_\alpha
(\tilde{g}_\Delta)_\beta+\tilde{g}_\beta
(\tilde{g}_\Delta)_\alpha] \eta ~,
\end{equation}
\begin{subequations}
\begin{multline}
\eta=\frac{m_{RR}}{64\pi^2}\left(\frac{M_3^2}{m_{RR}^2-M_3^2}
\ln\frac{m_{RR}^2}{M_3^2}-\frac{M_1^2}{m_{RR}^2-M_1^2}
\ln\frac{m_{RR}^2}{M_1^2}\right)\sin2\alpha_1
-\left[\left(\alpha_1,\,
M_1^2,\,M_3^2\right)\rightarrow\left(\alpha_2,\,
M_2^2,\,M_4^2\right)\right] ~,
\end{multline}
\begin{multline}
\tilde{\eta}=\frac{m_{RR}}{64\pi^2}\left(\frac{M_3^2}{m_{RR}^2-
M_3^2}\ln\frac{m_{RR}^2}{M_3^2}-\frac{M_1^2}{m_{RR}^2-M_1^2}
\ln\frac{m_{RR}^2}{M_1^2}\right)\sin2\alpha_1+\left[\left(\alpha_1,\,
M_1^2,\,M_3^2\right)\rightarrow\left(\alpha_2,\,
M_2^2,\,M_4^2\right)\right] ~.
\end{multline}
\end{subequations}
The parameters denoted by tilde are all suppressed by U(1)$_X$
 breaking terms.
Notice that for $\tilde{g}_\Delta=0$, the neutrino masses vanish.
This is expected as for $\tilde{g}_\Delta=0$, by assigning lepton
number equal to $+1$ and $-1$ respectively to $R$ and  $R^\prime$,
lepton number will be conserved so the neutrinos cannot have a
Majorana mass term. It is straightforward to confirm that
regardless of the flavor structure of the couplings, the
determinant of $m_\nu$ vanishes which means one of the neutrino
mass eigenvalues is zero and the neutrino mass scheme is
hierarchical. This structure is due to the fact that only two
right-handed neutrinos are incorporated within this model. In
order to make the neutrino mass scheme non-hierarchical ({\it
i.e.,} Det[$m_\nu]\ne 0$), another pair of $R$ and $R^\prime$
should be added.

  For $m_{RR}=300$ GeV and $m_\nu=0.05$ eV, it
has been found \cite{AMEND} that
\begin{subequations}
\begin{multline}
g \tilde{g}_\Delta \simeq  3.4 \times 10^{-6}
\frac{m_\nu}{0.05~{\rm eV}} \frac{70~{\rm GeV}}{M_1}\frac{50~{\rm
MeV}}{\delta}\frac{m_{RR}} {300~{\rm GeV}}\frac{0.1}{|\sin
\alpha_1|} \bigg(\frac{m_{RR}^2}{m_{RR}^2-m_\Delta^2}
\ln\frac{m_{RR}^2}{m_\Delta^2}+1
-\ln\frac{m_{RR}^2}{M_1^2}\bigg)^{-1} \\
  {\rm for}~~2\tilde{m}_\phi^2 m_{\phi \Delta}^2/m_\Delta^2 \simeq 2 M_1
  \delta |\sin \alpha_1|  \gg \tilde{m}_{\phi \Delta}^2 ~, \label{eq:numasscoup1}
\end{multline}
\begin{multline}
g \tilde{g}_\Delta \simeq
 3.3 \times 10^{-6}  \frac{m_\nu}{0.05~{\rm eV}}\frac{300~{\rm GeV}}{m_{RR}}\frac{1~{\rm GeV}^2}{\tilde{m}_{\phi\Delta}^2} \left(\frac{m_\Delta}{500~{\rm GeV}}\right)^2 \frac{m_{RR}^2-m_\Delta^2}{m_\Delta^2}\left(\ln
\frac{m_{RR}^2}{m_\Delta^2}\right)^{-1} \\
   {\rm for} ~~2\tilde{m}_\phi^2 m_{\phi
\Delta}^2/m_\Delta^2 \simeq 2 M_1 \delta |\sin \alpha_1|
 \label{eq:numasscoup2}\ll \tilde{m}_{\phi \Delta}^2 ~,
\end{multline}
\end{subequations}
\begin{equation}
\label{gTilde} \tilde{g} \tilde{g}_\Delta \simeq 1.3 \times
10^{-10}  \frac{m_\nu}{0.05~{\rm eV}}\frac{300~{\rm
GeV}}{m_{RR}}\frac{0.1}{|\sin
\alpha_1|}\frac{m_{RR}^2-m_\Delta^2}{m_\Delta^2}\left( \ln
\frac{m_{RR}^2}{m_\Delta^2}\right)^{-1} \ .
\end{equation}

The $g_\beta$, $\tilde{g}_\beta$ and $(\tilde{g}_\Delta)_\beta$
couplings will lead to LFV rare decays such as $l_\alpha \to
l_\beta \gamma$ \cite{AMEND}. Since $(\tilde{g}_\Delta)_\beta\ , \
\tilde{g}_\beta \ll g_\beta$, the dominant contribution is from
the $g$ coupling:
\begin{subequations}
\begin{align}
{\rm Br}(\mu \to e \gamma)&\approx  2.5 \times 10^{-9} \left(
\frac{300\GeV}{m_{RR}}\right)^4\left|\frac{g_\mu^*}{0.1}
\frac{g_e}{0.1} \right|^2 \quad \mathrm{and}\\
 {\rm Br}(\tau \to l_\alpha \gamma)&\approx 4.5\times 10^{-10}
\left( \frac{300\GeV}{m_{RR}}\right)^4\left|\frac{g_\tau^*}{0.1}
\frac{g_\alpha}{0.1}\right|^2 ~.
\end{align}
\end{subequations}
Let us now discuss the constraints on the parameters from bounds
$l_\alpha \to l_\beta \gamma$ \cite{pdg}. The bounds on
\mbox{Br$(\tau \to e\gamma)$} and \mbox{Br($\tau \to \mu \gamma$)}
allow  even values of $m_{RR}$ as small as 100~GeV and
$g_{\mu,\tau}$ as large as $0.2$. For $g_e, g_\mu \sim 0.1$, the
bound on Br($\mu \to e \gamma$) requires relatively large values
of $m_{RR}$, $m_{RR}\gtrsim1.1$~TeV. However, for $g_\mu\sim 0.02$
and $g_e\sim 0.01$, $m_{RR}$ as small as 100 GeV can  still be
accommodated.  An alternative solution is $g_e \ll g_\mu$ or $g_e
\gg g_\mu$. In the case $g_e \ll g_\mu$, the
$\tilde{g}\tilde{g}_\Delta \eta$ contribution dominates
$(m_\nu)_{e\alpha}$ ; {\it i.e.,}
$(m_\nu)_{e\alpha}=[\tilde{g}_e(\tilde{g}_\Delta)_\alpha+
\tilde{g}_\alpha(\tilde{g}_\Delta)_e]\eta$. Similarly for the case
$g_e \gg g_\mu$,
$(m_\nu)_{\mu\alpha}=[\tilde{g}_\mu(\tilde{g}_\Delta)_\alpha+
\tilde{g}_\alpha(\tilde{g}_\Delta)_\mu]\eta$.

\subsection{DM annihilation and searches for DM\label{DM}}

As discussed before, within this model the DM production is
thermal. To obtain the observed amount of the DM density, the
annihilation cross section should be
 \begin{equation}
 \langle \sigma
(\delta_1\delta_1 \to {\rm anything})v \rangle \simeq 3 \times
10^{-26}~{\rm cm}^3/{\rm sec} \ , \label{eq:orderOFcross}
\end{equation}
where $v$ is the relative velocity. In fact, since the mass
splitting between $\delta_1$ and $\delta_2$ might be small, the
effect of $\delta_2$ at the freeze-out epoch has to be taken into
account.

In \cite{AMEND}, different possible annihilation modes were
investigated. Depending on the regions of the parameter space,
different annihilation modes dominate. These modes are list as
follows: (1) Higgs mediated decay into $f \bar{f}$: $\delta_1
\delta_1 \to h^* \to f \bar{f}$; (2) Higgs mediated decay into an
on-shell $W$ plus and an off-shell $W$: $\delta_1 \delta_1 \to
WW^* \to W f \bar{f^\prime}$; (3) Annihilation into $WW$ or $WW^*$
via gauge interaction; (4) Annihilation into a Higgs pair. Of
course annihilation into $W$ or $H$ pair can be possible only for
heavy $M_1$. The annihilation of these particles in the sun center
can lead to a hard neutrino flux \cite{sunCENTER} which is
disfavored by the present indirect dark matter searches.
Annihilation modes to $\tau\bar{\tau}$ and $c\bar{c}$ are also
disfavored by indirect DM searches at the neutrino telescopes. We
shall focus in the range for which $m_b <M_1< 70~\GeV<m_W$. In
this range, the dominant annihilation mode is $\delta_1\delta_1
\to h^*\to b \bar{b}$ so the emerging neutrino flux from DM will
be rather soft and can be tolerated within the present bounds.

 The
coupling of $\delta_2$ and $\delta_1$ to the SM Higgs is given by
\begin{equation}\label{eq:hddcoupl}
\begin{split}
\lambda_Lv_H h \delta_i^2 \equiv
 \frac{ v_H}{2} \left( \big(  \lambda_{H\Delta 1} - \lambda_{H\Delta 2} \big) \sin^2 \alpha_1
+ \lambda_{H \phi} \cos^2 \alpha_1 - 2 \lambda_{H \Delta \phi}
\sin \alpha_1 \cos \alpha_1 \right)
h \delta_i^2\\
 = \frac{ \left( M_1^2 - \mu_\phi^2 \cos^2 \alpha_1 - \mu_\Delta^2
    \sin^2 \alpha_1 \right) }{v_H} h \delta_i^2
\end{split}
\end{equation}
where $i=1,2$ and the sub-dominant U(1)$_X$ violating terms are
neglected. Notice that the couplings of  $U(1)_X$ preserving part
of the Lagrangian can be made real by redefining the fields. In
particular, $\lambda_{H \Delta \phi}$ can be made real so there
will be no coupling of type $h \delta_1\delta_2$. The coupling of
form $h\delta_1\delta_2$ appears when both CP and $\U{1}_X$
symmetries are broken by {\it e.g.,} $\Im[\tilde{\lambda}_{H
\Delta \phi}]$.

The $\lambda_L$ coupling leads to
\begin{equation}
\label{eq:annihilationHiggs} \braket{\sigma(\delta_1\delta_1\to f
\bar f)_H v}=  N_c \frac{|\lambda_L|^2 }{\pi}
\frac{m_f^2}{(4\,M_1^2-m_h^2)^2} \frac{
(M_1^2-m_f^2)^{3/2}}{M_1^3} ~,
\end{equation}
  where $m_f$ is the fermion mass for the kinematically accessible
channels and $N_c=3 ~(1)$ for quarks (leptons). In the limit
$(M_2-M_1)/ (2 M_1)\ll 1$, we have to take into account the
annihilation of $\delta_2\delta_2$ in the calculation of the DM
abundance as
$$\braket{\sigma(\delta_2\delta_2\to f \bar f)_H v}\simeq \braket{\sigma(\delta_1\delta_1\to f
\bar f)_H v} ~. $$
Eventually $\delta_2$ decays via coupling to a virtual $Z$ boson
exchange \cite{AMEND}:
$$\Gamma(\delta_2 \to \delta_1 \nu \bar{\nu})\simeq 15
\left(\frac{M_2-M_1}{50\MeV}\right)^5\left(\frac{\sin\alpha_1}{0.1}\right)^4\,\mathrm{sec}^{-1}\;
.
$$
We therefore generally expect the decay to take place before Big
Bang Nucleosynthesis (BBN) so it should not affect the BBN
predictions.

The coupling in Eq.~(\ref{eq:hddcoupl}) leads to the interaction
of DM with nuclei via $t$-channel Higgs boson exchange so it can
be constrained by direct DM searches at underground experiments
sensitive to the recoil energy of the nuclei interacting with DM
particles. The cross section of the DM with nuclei can be written
as ~\cite{Andreas:2008xy}
\begin{equation}
\sigma_n \simeq \sigma_p=
\frac{|\lambda_L|^2}{\pi}\frac{\mu_{\delta_1n}^2
m_n^2}{M_1^2m_h^4} f^2\approx 6.5\times
10^{-45}\left(\frac{\lambda_L}{0.04}\right)^2\left(\frac{65~\GeV}{M_1}\right)^2\left(\frac{120\GeV}{m_h}\right)^4
\left(\frac{f}{0.2}\right)^2\mathrm{cm}^2 ~, \label{eSI}
\end{equation}
where $\mu_{\delta_1n}$ is the reduced mass of the dark
matter-neutron system, $m_n$ is the nucleon mass and $f$ is the
nuclear matrix element parameter which can vary within the present
uncertainties, $0.14 < f < 0.66$ ~\cite{Andreas:2008xy}. The same
coupling also determines annihilation of DM so we can write
\begin{equation}
\sigma_n \simeq\sigma_p= \frac{f^2\mu_{\delta_1n}^2m_p^2
(4M_1^2-m_h^2)^2}{\pi
M_1m_h^4v_H^2}\frac{\braket{\sigma(\delta_1\delta_1\to
h^*\to\mathrm{SM\,final\,states})v}}
{4\Gamma(h\to\mathrm{SM\,final\,states})|_{m_h\to2 M_1}} ~.
\label{sigmansigmaann}
\end{equation}
The strong bound on \cite{xenonPRL} already constrains a
considerable part of the parameter space. We generally expect a
signal at future DM search experiments unless $M_1 \to m_h/2$. The
case of $M_1\to m_h/2$ seems to be unnatural. If the future
searches for DM do not find a signal and no neutral stable scalar
with mass equal to $m_h/2$ is found by colliders, this model will
be disfavored and will eventually be ruled out.
\subsection{Electroweak precision tests\label{precision}}
The new particles added to SM participate in the electroweak
interactions so they can lead to  corrections to the electroweak
precision parameters. The contributions to the oblique parameters
$\hat S,\, \hat T,\, W,\,Y$~\cite{Barbieri} were explicitly
calculated in \cite{AMEND}.

Since $R$ and $R^\prime$ have equal masses, their contributions to
$\hat S$ and $\hat T$ parameter exactly cancel each other. Their
contribution to  $W$ and $Y$ will be non-zero but tiny:
\begin{equation*}
W=\frac{g_{\SU{2}}^2}{120\pi^2}\frac{m_W^2}{m_{RR}^2}\quad\quad\mathrm{and}\quad\quad
Y=\frac{g_{\U{1}}^2}{120\pi^2}\frac{m_W^2}{m_{RR}^2}\; .
\end{equation*}

 The contribution of $\Delta$ to the oblique parameters can be written as
\begin{align}
\hat S&=\frac{g_{\SU{2}}^2}{24 \pi^2} \xi\,,& \hat T&=\frac{25\,
g_{\SU{2}}^2}{576\pi^2}\frac{m_\Delta^2}{m_W^2} \xi^2\,,&
W&=-\frac{7\, g_{\SU{2}}^2}{720\pi^2}\frac{m_W^2}{m_\Delta^2}\,,&
Y&=-\frac{7\, g_{\U{1}}^2}{480\pi^2} \frac{m_W^2}{m_\Delta^2} ~,
\end{align}
where the relation
$2\,m_{\Delta^{+}}^2=m_{\Delta}^2+m_{\Delta^{++}}^2$ has been used
and the results have been expanded in
\begin{equation}
\xi\equiv\frac{m_{\Delta^{++}}^2-m_\Delta^2}{m_\Delta^2}=\lambda_{H\Delta2}\frac{v_H^2}{m_\Delta^2}\;.
\end{equation}
The contribution of $\phi$ to the electroweak precision parameters
 is suppressed by a factor of $|\sin \alpha_1 \sin \alpha_2|$
relative to that of $\Delta$ so it can be neglected. Within this
model, Higgs can be heavier than in the SM model because the new
contributions can cancel out the contributions from the Higgs to
the oblique parameters and the upper bounds from electroweak
precision tests on the Higgs mass can be relaxed. Without
cancelation ({\it i.e.,} for a light Higgs mass), the $\hat T$
parameter constrains $\xi\lesssim 0.1$ which translates into a
bound on the splitting of the components of the triplet. This
results in a mild bound on $\lambda_{H\Delta2}$, {\it e.g.}, for
$m_\Delta\simeq 500\GeV$, the bound is $\lambda_{H\Delta2}\lesssim
0.5$.
\subsection{Signatures at colliders \label{collider}}
We expect a rich phenomenology for the LHC within this model. For
$M_1,M_2<m_h/2$, the SM Higgs boson can decay into a pair of
$\delta_1$ or  a pair of $\delta_2$. These new decay modes can
dominate over the decay to the $b\bar{b}$ pair when $\lambda_L
\stackrel{>}{\sim}m_b/v_H\simeq 0.02$. Decay mode to $\delta_1$
pair will appear as missing energy. If $M_2-M_1$ is larger than
$2m_e$, $\delta_2$ can also decay into $\delta_1 e^-e^+$ which
appears as a distinct displaced vertex. In order for the decay to
take place within the detector, the following condition is
necessary:
 $d\,
\Gamma_{\delta_2}/2\gamma \gtrsim v$ where $v$ is the velocity of
$\delta_2$ and $\gamma=(1-v^2)^{-1/2}$. $d$ characterizes the size
of the detector. As shown in \cite{AMEND} this condition requires
$|M_2-M_1|>500$~MeV. Remember that $M_2-M_1$ is suppressed by the
$\U{1}_X$ violating parameters. For smaller values of the
splitting the decay mode $H\to \delta_2\delta_2$ will also appear
as missing energy signal regardless of if the $\delta_2\to
\delta_1 e^-e^+$ mode is kinematically accessible or not.

If the new  particles are not too heavy, the charged particles
$\Delta^{++}$, $\Delta^+$, $E_R^-$ and $E_R'^+$ as well as the
neutral particles $\delta_3$, $\delta_4$, $\nu_R$ and $\nu_R'$ can
be produced through electroweak interactions.
 They will eventually decay
into the SM particles plus $\delta_1$ or $\delta_2$. In particular
$\Delta^{++}$ can decay to a pair of same-sign charged leptons:
$$\Delta^{++} \to l_\alpha^+ l_\beta^+\delta_{1,2} \ .$$
The background from the SM to the same sign charged lepton pair
signal is not very high which makes the discovery of $\Delta^{++}$
easier. $\Gamma(\Delta^{++}\to \ell_\alpha^+ \ell_\beta^+
\delta_{1,2})$ is proportional to $|(\tilde{g}_\Delta)_\alpha
g_\beta+(\tilde{g}_\Delta)_\beta g_\alpha|^2$. The decays of the
charged components of $R$ and $R^\prime$ are given by the
$g_\alpha$ couplings: ${\rm Br}(E_R^-\to
\ell_\alpha^-\delta_{1,2})\propto |g_\alpha|^2$. In principle,
$g_\alpha$ and $(\tilde{g}_\Delta)_\alpha$ can be directly
extracted by studying the flavor pattern of the decay modes of
$\Delta^{++}$ and $E_R$.  Notice that the coupling determining
${\rm Br}(E_R^-\to \ell_\alpha^-\delta_{1,2})$ also determines $
{\rm Br}(l_\alpha^- \to l_\beta^-\gamma)$. Moreover the
combination $(\tilde{g}_\Delta)_\alpha
g_\beta+(\tilde{g}_\Delta)_\beta g_\alpha$ determining
$\Gamma(\Delta^{++}\to \ell_\alpha^+ \ell_\beta^+ \delta_{1,2})$
is exactly the combination appearing in the neutrino mass matrix
in Eq.~(\ref{mmmnnnn}).
\section{Concluding remarks \label{con}}
In this letter, we have first reviewed the SLIM scenario
\cite{Boehm} and the model embedding it \cite{myModel}. We have
then reviewed  AMEND, which stands for A Model Explaining Neutrino
masses and Dark matter  \cite{AMEND}. Both these models are
constructed to explain the tiny neutrino mass and provide us with
a DM candidate. We have reviewed the implications of these models
in various experiments and observations. We have updated the
results in \cite{Boehm,myModel,AMEND}, taking into account the
most recent data release from experiments such as direct DM
searches XENON10 \cite{XENON10} and XENON100 \cite{XENON100}.

In both models, there is a $Z_2$ symmetry that makes DM stable and
forbids a Dirac mass term for neutrinos. The neutrinos acquire
Majorana masses at one loop level. In the framework of the both
models, we generally expect the value of Br($\mu \to e \gamma$) to
be within the reach of MEG. The present bound already rules out a
part of the parameter space.

 The SLIM model has a light
sector (mass$<10$~MeV) and a heavy sector (mass$>m_W$). The light
sector consists of the scalar DM candidate, $\delta_1$ and at
least one right-handed neutrino, $N_1$. The heavy sector consists
of the components of a scalar electroweak doublet: (i) a CP-odd
neutral scalar ($\phi_2$); (ii)  a CP-even neutral scalar
($\delta_2$); (iii) a charged scalar ($\phi^-$). The light sector
can show up in supernova explosion and the decay of light mesons.
The present bounds from meson decay as well as supernova explosion
are too weak to constrain the model. However, the on-going KLOE
experiment \cite{Kloe} and future NA62 experiment \cite{NA62} can
test the model. In case of future observation of supernova
explosion, invaluable information on this bound can be derived.
Considering the fact that there is a lower bound on the coupling
of the light sector to active neutrinos and an upper bound on the
masses of the light sector, this model is falsifiable by low
energy  experiments with enough luminosity.

The heavy sector can be produced at the LHC via the electroweak
interactions. The present lower bounds are $m_{\phi^-}>80$ GeV
\cite{chargedHiggs} and $m_{\delta_2},m_{\phi_2}>90$~GeV
\cite{neutralHiggs}. For relatively light new particles
($m_{\phi^-}<130$ GeV), LHC with $30~{\rm fb}^{-1}$ of integrated
luminosity and at 14 TeV center of mass energy can make discovery
via the $pp\to \phi^+\phi^-\to \mu^-\mu^++{\rm missing~ energy}$
mode. In principle, by studying $\phi^- \to l_\alpha^-+{\rm
missing~energy}$, the coupling $|g_{1 \alpha}|^2$ (see
Eq.~(\ref{Lag})) or for the case that $m_{N_2}\ll m_{\phi^-}$,
$|g_{1 \alpha}|^2+|g_{2 \alpha}|^2$ can be extracted. These are
the same couplings that determine the neutrino mass matrix and the
pattern of LFV rare decays. However, as discussed in \cite{AMEND},
to make this possible the large uncertainty induced by the
uncertainties in the parton distribution functions should be
reduced or the energy of center of mass should be increased to
enhance the signal to background ratio.

The other model that we discussed (AMEND) does not necessarily
contain a low energy sector. Within this model, a scalar
electroweak triplet ($\Delta$) and two fermionic doublets with
opposite hypercharges exist. $\Delta$ contains a doubly charged
component, $\Delta^{++}$. This particle can lead to  signals
consisting of a pair of same sign charged leptons plus missing
energy. The background to this signal from the SM is not very high
so the discovery chance of the signal should be high provided that
$\Delta^{++}$ is not too heavy.

These two models have quite different predictions for direct DM
searches. Within the SLIM model, we do not expect a signal in the
future DM searches. However, within AMEND, we generally expect a
signal in the future DM searches. Only in some very specific parts
of the parameter space such as when the DM mass approaches half
the Higgs mass, the DM nucleon cross section is suppressed. The
present bounds from direct DM searches already rule out a part of
the parameter space.
\section*{Acknowledgements}
I would like to thank C. Boehm, T. Hambye, M. Hashemi, S.
Palomares-Ruiz, S. Pascoli and M. A. Schmidt who were my
collaborators in the projects upon which this proceeding is based.
I would also like to thank G. Colangelo and E. Ma for useful
discussions. I am especially grateful for the organizers of
International Conference on Flavor Physics in the LHC Era held in
Singapore for their hospitality.


\begin{thebibliography}{99}
\bibitem{MEG}
  G.~Cavoto,
  arXiv:1012.2110 [hep-ex].


\bibitem{Boehm}
  C.~Boehm, Y.~Farzan, T.~Hambye, S.~Palomares-Ruiz and S.~Pascoli,
  Phys.\ Rev.\  D {\bf 77}, 043516 (2008)
  [arXiv:hep-ph/0612228].
   \bibitem{myModel}
  Y.~Farzan,
  Phys.\ Rev.\  D {\bf 80}, 073009 (2009)
  [arXiv:0908.3729 [hep-ph]].
\bibitem{AMEND}
Y.~Farzan, S.~Pascoli, M.~A.~Schmidt,
  JHEP {\bf 1010}, 111 (2010).
  [arXiv:1005.5323 [hep-ph]].
\bibitem{serpicoraffelt}
  P.~D.~Serpico and G.~G.~Raffelt,
  Phys.\ Rev.\  D {\bf 70}, 043526 (2004)
  [arXiv:astro-ph/0403417].

 \bibitem{silvia}
   S.~Palomares-Ruiz and S.~Pascoli,
  Phys.\ Rev.\  D {\bf 77}, 025025 (2008)
  [arXiv:0710.5420 [astro-ph]].
\bibitem{LENA}
  T.~Marrodan Undagoitia {\it et al.},
  J.\ Phys.\ Conf.\ Ser.\  {\bf 120}, 052018 (2008).
  \bibitem{Farzan}
  Y.~Farzan,
  Mod.\ Phys.\ Lett.\  A {\bf 25}, 2111 (2010)
  [arXiv:1009.1234 [hep-ph]].
 \bibitem{Britton:1993cj}
  D.~I.~Britton {\it et al.},
  Phys.\ Rev.\  D {\bf 49}, 28 (1994).

\bibitem{Barger:1981vd}
  V.~D.~Barger, W.~Y.~Keung and S.~Pakvasa,
  Phys.\ Rev.\  D {\bf 25}, 907 (1982).

\bibitem{Gelmini:1982rr}
  G.~B.~Gelmini, S.~Nussinov and M.~Roncadelli,
  Nucl.\ Phys.\  B {\bf 209}, 157 (1982).

\bibitem{Lessa}
  A.~P.~Lessa and O.~L.~G.~Peres,
  Phys.\ Rev.\  D {\bf 75}, 094001 (2007)
  [arXiv:hep-ph/0701068].
  \bibitem{Kloe}
  F.~Ambrosino {\it et al.}  [KLOE Collaboration],
  Eur.\ Phys.\ J.\  C {\bf 64}, 627 (2009)
  [Erratum-ibid.\  {\bf 65}, 703 (2010)]
  [arXiv:0907.3594 [hep-ex]].
\bibitem{73}
  C.~Y.~Pang, R.~H.~Hildebrand, G.~D.~Cable and R.~Stiening,
  Phys.\ Rev.\  D {\bf 8}, 1989 (1973).
  \bibitem{NA62}
  http://na62.web.cern.ch/NA62/Home/Home.html
\bibitem{G-H}
H.~E.~Haber,
  Phys.\ Rev.\  D {\bf 67} (2003) 075019
  [arXiv:hep-ph/0207010].








\bibitem{chargedHiggs}

    [LEP Higgs Working Group for Higgs boson searches and ALEPH Collaboration
                  an],
  arXiv:hep-ex/0107031;
G.~Abbiendi {\it et al.}  [OPAL Collaboration],
  Eur.\ Phys.\ J.\  C {\bf 32} (2004) 453
  [arXiv:hep-ex/0309014];
G.~Abbiendi {\it et al.}  [OPAL Collaboration],
  arXiv:0812.0267 [hep-ex].

\bibitem{neutralHiggs}
\url{http://lephiggs.web.cern.ch/LEPHIGGS/papers/July2005_MSSM/LHWG-Note-2005-01.pdf}.
\bibitem{inert}
  V.~I.~Kuvshinov, V.~I.~Kashkan and R.~G.~Shulyakovsky,
  arXiv:hep-ph/0107031.
\bibitem{stefano}
S.~Profumo,
  Phys.\ Rev.\  D {\bf 78}, 023507 (2008)
  [arXiv:0806.2150 [hep-ph]].
\bibitem{bullet}
 S.~W.~Randall, M.~Markevitch, D.~Clowe, A.~H.~Gonzalez and M.~Bradac,
  arXiv:0704.0261 [astro-ph].
\bibitem{Lavoura}
  L.~Lavoura,
  Eur.\ Phys.\ J.\  C {\bf 29}, 191 (2003)
  [arXiv:hep-ph/0302221].
\bibitem{PDG}
 K. Nakamura et al. (Particle Data Group), J. Phys. G 37, 075021 (2010).
\bibitem{Hashemi}
Y.~Farzan and M.~Hashemi,
  JHEP {\bf 1011}, 029 (2010)
  [arXiv:1009.0829 [hep-ph]].

\bibitem{Dama}
  R.~Bernabei {\it et al.},
  Eur.\ Phys.\ J.\  C {\bf 67}, 39 (2010)
  [arXiv:1002.1028 [astro-ph.GA]].

\bibitem{XENON10}
  J.~Angle {\it et al.}  [XENON10 Collaboration],
  arXiv:1104.3088 [astro-ph.CO].


\bibitem{TuckerSmith}
  D.~Tucker-Smith, N.~Weiner,
  Phys.\ Rev.\  {\bf D64}, 043502 (2001).
  [hep-ph/0101138].
\bibitem{lightDM}

 F.~Petriello and K.~M.~Zurek,
  JHEP {\bf 0809}, 047 (2008)
  [arXiv:0806.3989 [hep-ph]].
\bibitem{XENON100}
  E.~Aprile {\it et al.}  [XENON100 Collaboration],
  arXiv:1104.3121 [astro-ph.CO].

\bibitem{sunCENTER}
M.~Cirelli, N.~Fornengo, T.~Montaruli, I.~A.~Sokalski, A.~Strumia,
F.~Vissani,
  Nucl.\ Phys.\  {\bf B727}, 99-138 (2005);
 A.~E.~Erkoca, M.~H.~Reno, I.~Sarcevic,
  Phys.\ Rev.\  {\bf D80}, 043514 (2009).
  [arXiv:0906.4364 [hep-ph]].
\bibitem{Andreas:2008xy}
  S.~Andreas, T.~Hambye and M.~H.~G.~Tytgat,
  JCAP {\bf 0810}, 034 (2008)
  [arXiv:0808.0255 [hep-ph]].
\bibitem{xenonPRL}
  E.~Aprile {\it et al.} [ XENON100 Collaboration ],
  [arXiv:1104.2549 [astro-ph.CO]].
\bibitem{Barbieri}
R.~Barbieri, A.~Pomarol, R.~Rattazzi, A.~Strumia,
  Nucl.\ Phys.\  {\bf B703}, 127-146 (2004);
  G.~Cacciapaglia, C.~Csaki, G.~Marandella, A.~Strumia,
  Phys.\ Rev.\  {\bf D74}, 033011 (2006).
  [hep-ph/0604111].



\end{thebibliography}
\end{document}